\documentclass{article}

\usepackage[preprint]{nips_2018}

\makeatletter
\newcommand\footnoteref[1]{\protected@xdef\@thefnmark{\ref{#1}}\@footnotemark}
\makeatother

\usepackage{placeins}
\usepackage{xcolor}
\PassOptionsToPackage{hyphens}{url}
\usepackage[hidelinks]{hyperref}
\usepackage{footmisc}
\usepackage{multirow}
\usepackage{tikz-qtree}
\usepackage{amsmath}

\usepackage{url}
\usepackage{breakurl}
\defcitealias{ftc-collab}{(FTC/DoJ, 2000)}

\tikzset{
  solid node/.style={circle,draw,inner sep=2,fill=black},
  hollow node/.style={circle,draw,inner sep=2},
  empty node/.style={rectangle,draw,fill=white,color=white}
}

\usepackage[utf8]{inputenc} 
\usepackage[T1]{fontenc}    
\usepackage{url}            
\usepackage{hyperref}
\usepackage{booktabs}       
\usepackage{nicefrac}       
\usepackage{microtype}      
\usepackage{amsfonts}
\usepackage{amsmath}
\usepackage{amssymb}
\usepackage{wrapfig}
\usepackage{mathtools}
\usepackage{subcaption}
\usepackage{verbatim}
\usepackage[capitalize,nameinlink]{cleveref}
\usepackage[inline]{enumitem}   
\usepackage{makecell, tabularx}
\usepackage{floatpag}

\newif\ifanony  
\anonyfalse

\usepackage{natbib}
\bibpunct{(}{)}{;}{a}{,}{,}

\AtBeginDocument{

}

\title{The Role of Cooperation in \\ \vspace{0.2cm} Responsible AI Development}
\author {Amanda Askell\thanks{Primary/corresponding author.}\hspace{1.3mm}\\ 
  OpenAI \\ 
  \texttt{amanda@openai.com} \\
  \And
  Miles Brundage \\
  OpenAI \\ 
  \texttt{miles@openai.com} \\
  \And
  Gillian Hadfield \\
  OpenAI \\ 
  \texttt{gillian@openai.com}}

\date{\today}

\begin{document}

\maketitle

\begin{abstract}

In this paper, we argue that competitive pressures could incentivize AI companies to underinvest in ensuring their systems are safe, secure, and have a positive social impact. Ensuring that AI systems are developed responsibly may therefore require preventing and solving collective action problems between companies.
We note that there are several key factors that improve the prospects for cooperation in collective action problems. We use this to identify strategies to improve the prospects for industry cooperation on the responsible development of AI.

\end{abstract}

\section*{Introduction} 
\setcounter{footnote}{0}

Machine learning (ML) is used to develop increasingly capable systems targeted at tasks like voice recognition, fraud detection, and the automation of vehicles. These systems are sometimes referred to as narrow artificial intelligence (AI) systems. Some companies are also using machine learning techniques to try to develop more general systems that can learn effectively across a variety of domains rather than in a single target domain. 
Although there is a great deal of uncertainty about the development path of future AI systems---whether they will remain specialized or grow increasingly general, for example---many agree that if the current rate of progress in these domains continues then it is likely that advanced artificial intelligence systems will have an increasingly large impact on society.  

This paper focuses on the private development of AI systems that could have significant expected social or economic impact, and the incentives AI companies have to develop these systems responsibly. Responsible development involves ensuring that AI systems are safe, secure, and socially beneficial. 

In most industries, private companies have incentives to invest in developing their products responsibly. These include market incentives, liability laws, and regulation.
We argue that AI companies have the same incentives to develop AI systems responsibly, although they appear to be weaker than they are in other industries. 
Competition between AI companies could decrease the incentives of each company to develop responsibly by increasing their incentives to develop faster. As a result, if AI companies would prefer to develop AI systems with risk levels that are closer to what is socially optimal---as we believe many do---responsible AI development can be seen as a collective action problem.\footnote{AI research companies increasingly have teams dedicated to the safe and ethical development of technology and many large technology companies participate in voluntary efforts to articulate and establish principles and guidelines, and in some cases call for government regulation, to address AI-related risks.}

We identify five key factors that make it more likely that companies will be able to overcome this collective action problem and cooperate---develop AI responsibly with the understanding that others will do likewise. These factors are: high trust between developers (\textit{High Trust}), high shared gains from mutual cooperation (\textit{Shared Upside}), limited exposure to potential losses in the event of unreciprocated cooperation (\textit{Low Exposure}), limited gains from not reciprocating the cooperation of others (\textit{Low Advantage}), and high shared losses from mutual defection (\textit{Shared Downside}). 

Using these five factors, we identify four strategies that AI companies and other relevant parties could use to increase the prospects for cooperation around responsible AI development. These include correcting harmful misconceptions about AI development, collaborating on shared research and engineering challenges, opening up more aspects of AI development to appropriate oversight, and incentivizing greater adherence to ethical and safety standards. This list is not intended to be exhaustive, but to show that it is possible to take useful steps towards more responsible AI development.

The paper is composed of three sections. In \hyperref[section1]{section 1}, we outline responsible AI development and its associated costs and benefits. In \hyperref[section2]{section 2}, we show that competitive pressures can generate incentives for AI companies to invest less in responsible development than they would in the absence of competition, and outline the five factors that can help solve such collective action problems. In \hyperref[sec:factors]{section 3}, we outline the strategies that can help companies realize the gains from cooperation. We close with some questions for further research.

\section{The benefits and costs of responsible AI development}\label{section1}

AI systems have the ability to harm or create value for the companies that develop them, the people that use them, and members of the public who are affected by their use. 
In order to have high expected value for users and society, AI systems must be safe---they must reliably work as intended---and secure---they must have limited potential for misuse or subversion. AI systems should also not introduce what \cite{risktypes} call ``structural risks'', which involve shaping the broader environment in subtle but harmful ways.\footnote{\cite{risktypes} argue that the ``the accident-misuse dichotomy obscures how technologies, including AI, often create risk by shaping the environment and incentives''. We restrict accident risks to technical accidents and misuse risks to direct misapplications of a system. `Structural risks', as we use the term, are intended to capture the broader impact of AI systems on society and social institutions.} The greater the harm that can result from safety failures, misuse, or structural risks, the more important it is that the system is safe and beneficial in a wide range of possible conditions \citep{dunn2003designing}. This requires the responsible development of AI. 

\subsection{What is responsible AI development?}

AI systems are increasingly used to accomplish a wide range of tasks, some of which are critical to users' health and wellbeing. As the range of such tasks grows, the potential for accidents and misuse also grows, raising serious safety and security concerns \citep*{amodei2016concrete, 2018arXiv180207228B}. Harmful scenarios associated with insufficiently cautious AI development have already surfaced with, for example, biases learned from large datasets distorting decisions in credit markets and the criminal justice system, facial recognition technologies disrupting established expectations of privacy and autonomy, and auto-pilot functions in some automobiles causing new types of driving risk (while reducing others). Longer term, larger scale scenarios include dangers such as inadvertent escalation of military conflict involving autonomous weapon systems or widespread job displacement.

Responsible AI development involves taking steps to ensure that AI systems have an acceptably low risk of harming their users or society and, ideally, to increase their likelihood of being socially beneficial. This involves testing the safety and security of systems during development, evaluating the potential social impact of the systems prior to release, being willing to abandon research projects that fail to meet a high bar of safety, and being willing to delay the release of a system until it has been established that it does not pose a risk to consumers or the public. Responsible AI development comes in degrees but it will be useful to treat it as a binary concept for the purposes of this paper. We will say that an AI system has been developed responsibly if the risks of it causing harms are at levels most people would consider tolerable, taking into account their severity, and that the amount of evidence grounding these risk estimates would also be considered acceptable.\footnote{This will generally mean if an AI system is developed responsibly, the risk of irreversible catastrophic harm from that system---whether through accident, misuse, or negative social impact---must be very low. This is consistent with what \cite{sunstein2005irreversible} calls the `Irreversible Harm Precautionary Principle'.}  

Responsible AI development involves work on safety, security, and the structural risks associated with AI systems. Work on the safety of AI aims to mitigate accident risks \citep*{amodei2016concrete} and ensure that AI systems function as intended \citep*{Building2:online} and behave in ways that people want \citep{irving2018ai}. Work on the security of AI aims to prevent AI systems from being attacked, co-opted, or misused by bad actors \citep*{2018arXiv180207228B}.\footnote{Mitigating misuse risks is sometimes included under AI safety, broadly construed \citep{AIsafet70:online}} Work evaluating the structural impact of AI aims to identify and mitigate both the immediate and long term structural risks that AI systems pose to society: risks that don't quite fit under narrow definitions of accident and misuse. These include joblesness,  military conflict, and threats to political and social institutions.\footnote{The literature on the societal impact of AI is vast. See \cite{cummings2017artificial} on AI and warfare, for example. For a broader overview see \cite{dafoe2018ai}.}

\subsection{The cost of responsible AI development}

It is likely that responsible development will come at some cost to companies, and this cost may not be recouped in the long-term via increased sales or the avoidance of litigation. In order to build AI systems responsibly, companies will likely need to invest resources into data collection and curation, system testing, research into the possible social impacts of their system, and, in some cases, technical research to guarantee that the system is reliably safe. In general, the safer that a company wants a product to be, the more constraints there are on the kind of product the company can build and the more resources it will need to invest in research and testing during and after its development.

If the additional resources invested in ensuring that an AI system is safe and beneficial could have been put towards developing an AI system with fewer constraints more quickly, we should expect responsible AI development to require more time and money than incautious AI development. This means that responsible development is particularly costly to companies if the value of being the first to develop and deploy a given type of AI system is high (even if the first system developed and deployed is not demonstrably safe and beneficial). 

There are generally several advantages that are conferred on the first company to develop a given technology \citep{lieberman1988first}. If innovations can be patented or kept secret, the company can gain a larger share of the market by continuing to produce a superior product and by creating switching costs for users. Being a first-mover also allows the company to acquire scarce resources ahead of competitors. 
If hardware, data, or research talent become scarce, for example, then gaining access to them early confers an advantage.\footnote{As \cite{10.2307/2118212} notes, larger first-movers can also spread their prior investment in R\&D over a larger number of applications.} 
And if late movers are not able to catch up quickly then first-mover advantages will be greater.
In the context of AI development, having a lead in the development of a certain class of AI systems could confer a first mover advantage. This effect would be especially pronounced in the case of discontinuous changes in AI capabilities\footnote{\label{fn-discont} The possibility of discontinuous progress in AI is discussed by \cite{good1966speculations}, \cite{chalmers2009singularity}, \cite{Yudkowsky_intelligenceexplosion}, \cite{shanahan2015technological} and \cite{bostrom2017superintelligence}. \cite{Likeliho38:online} provide a critical overview of the arguments for discontinuity and \cite{Takeoffs13:online} presents arguments against the claim.}, but such a discontinuity is not necessary in order for a first mover advantage to occur. 

Responsible development may therefore be costly both in terms of immediate resources required, and in the potential loss of a first-mover advantage. Other potential costs of responsible AI development include performance costs and a loss of revenue from not building certain lucrative AI systems on the grounds of safety, security, or impact evaluation. An example of a performance cost is imposing a limit on the speed that self-driving vehicles can travel in order to make them safer. An example of revenue loss is refusing to build a certain kind of facial recognition system because it may undermine basic civil liberties \citep{Facialre69:online}.

AI companies may not strongly value being the first to develop a particular AI system because first-mover advantages do not always exist. Indeed, there are often advantages to entering a market after the front-runner. These include being able to free-ride on the R\&D of the front-runner, to act on more information about the relevant market, to act under more regulatory certainty, and having more flexible assets and structures that let a company respond more effectively to changes in the environment \citep{gilbert1996innovation}. These can outweigh the advantages of being the first to enter that same market. And it has been argued that late mover advantages often do outweigh first mover advantages \citep{markides2004fast, querbes2017evolving}. We therefore acknowledge that the assumption that there will be a first-mover advantage in AI development may not be true. If a first-mover advantage in AI is weak or non-existent then companies are less likely to engage in a race to the bottom on safety since speed is of lower value. Instead of offering predictions, this paper should be thought of as an analysis of more pessimistic scenarios that involve at least a moderate first mover advantage.

Much of the discussion of AI development races assumes that they have a definitive endpoint. Although some have hypothesized that if AI progress is discontinuous or  sufficiently rapid then it could essentially have a definitive endpoint, the case for this remains speculative.\footnote{If AI development is extremely rapid then gaps between each company would likely increase over time. This means that a company that is ahead of others may at some point be ahead of them by a great deal in strategically important areas, and could use this to undermine their competitors \cite[pp. 91-104]{bostrom2017superintelligence}} It is therefore important to note that AI development may take the form of a perpetual R\&D race: a race to stay technologically ahead of competitors rather than a race to reach some particular technological endpoint \citep{aoki1991r, breitmoser2010understanding}.
If this is the case then AI companies would still have an incentive to speed up development in order to stay ahead of others, especially if the gap between companies was small.
The present analysis is applicable to perpetual races in which there is at least a moderate first mover advantage, several companies are competing to stay ahead, and leadership is not yet entrenched.\footnote{\cite{breitmoser2010understanding} note that perpetual R\&D races tend to collapse into leadership monopolies. The larger the gap between the front-runner and the company in second place in a perpetual race, the less of an incentive the front-runner has to trade safety for speed.}

\subsection{The benefits of responsible AI development}

In the law and economics literature on product safety, it is generally accepted that market forces create incentives for companies to invest in making their products safe \citep{oi1973economics}. Suppose that companies have accurate information about how safe the products they are developing are and that consumers have access to accurate information about how safe a company's product is, either prior to release or by observing the harms caused by a product after it is released \citep{ben1998should, chen2017competition}.\footnote{See \cite{daughety2013economic}, who make similar assumptions in their idealized model of markets.} If consumers have a preference for safer products and respond rationally to this preference, they will not buy products that are insufficiently safe, or will pay less for them than for safer alternatives \citep{polinsky2009uneasy}. Releasing unsafe products will also result in a costly loss of reputation for companies \citep{daughety1995product}.\footnote{\cite{rhee2006liability} provide evidence that the relationship between safety failures and reputation loss may be more complex than this, however.} Finally, releasing unsafe products could result in burdensome regulation of the industry or in litigation costs. Therefore companies that are concerned about a sufficiently long time-horizon involving repeated interaction with customers, regulators, and other stakeholders that incentivize safety should internalize the value of responsible development.

Market forces alone may not always incentivize companies to invest the appropriate amount into ensuring their products are safe. If consumers cannot get access to information about the safety of a product---how likely safety failures are or how costly they are---then companies have an incentive to under-invest in safety. And if companies have inaccurate information about the safety of the products they are developing, they will not invest in safety to the degree demanded by consumers. Finally, poor corporate governance can result in suboptimal decisions about risk \citep{cai2010compensation}. Product liability law and safety regulation are intended to correct such market failures by providing consumers with information about products, incentivizing companies to invest more in safety, and compensating consumers that are harmed by product safety failures \citep{hylton2012law, 10.2307/724257}.\footnote{See \cite{stiglitz2009regulation} on government regulation as a response to market failures or inefficiencies, but note that actual motivations for government regulation are typically more complicated \citep{henson1999food}.  \cite{calabresi1970costs} provides comprehensive overview of of the role of law in the minimization of costs from safety failures. The relationship between product liability law and safety regulations---in particular, whether it is efficient to use them jointly---is a matter of some debate \citep{shavell1984liability, 10.2307/2006714}.}

We may expect companies to under-invest in safety if the costs to consumers don't result in commensurate costs for the company; either via a reduction in revenue, reputation loss, fines from regulators, or successful litigation by consumers. Safety failures can also affect those who do not consume the product, however. Consider a 2018 recall of over 8,000 Volkswagen vehicles potentially affected by a brake caliper issue that could result in increased stopping distances or loss of vehicle control \citep{Volkswag27:online}. A safety failure resulting from this could harm not only the vehicle's occupants but also pedestrians and other drivers.\footnote{\label{fn-liability}There is currently uncertainty about who should be held liable for the harms that the safety failures of autonomous systems inflict on the public \citep{schellekens2015self,WhoIsLia40:online}.} Harms that safety failures inflict on non-consumers are negative externalities, and benefits that safer products produce for non-consumers are positive externalities. We should anticipate companies under-investing in reducing negative externalities and increasing positive externalities relative to their social value, since the costs and benefits this produces for society don't result in commensurate costs and benefits for the company \citep{dahlman1979problem}. 

To give a concrete example, consider facial recognition technology. Microsoft have argued that this technology could be used in ways that many would consider harmful: to violate individuals' privacy or suppress their political speech, for example \citep{Facialre62:online, Facialre69:online}. Even if companies would prefer to build facial recognition systems that cannot be misused, either to avoid causing harm or to avoid the reputation costs of this harm, the cost of developing safeguards may not outweigh their benefits if companies cannot be held liable for these harms and there is no regulation preventing misuse. For this reason, Microsoft has called for regulation that would require that companies invest in measures that reduce the risks from facial recognition technology, and that could also mitigate potential misuse of the technology by commercial entities or by governments \citep{Facialre62:online}.

The discussion thus far treats companies as though they were motivated only by profit, i.e. they only care about things like reputation and product safety insofar as they are a means to make more profit or avoid losses. This view is common in the literature on corporate social responsibility \citep{Campbell2007, devinney2009socially} but it is clearly an abstraction. Companies are run by, invested in, and composed of humans that care about the impact their products will have on the world and on other people. Employees at technology companies have already shown that they care a great deal about the social implications of the systems they are building \citep{TheNewfo81:online}. 

 The things that motivate AI companies other than profits, such as benefiting people rather than harming them, will generally push even more in favor of responsible development: they will rarely push against it. Assuming that companies are motivated solely by profit therefore lets us analyze a kind of `worst case scenario' for responsible development. We will therefore often treat companies as though they were driven solely by profit, even though we do not find this plausible. It is important that the reader bear this in mind, since treating companies as profit-driven entities can be self-fulfilling, and can therefore contribute to the very problems we are attempting to solve.

\subsection{Are existing incentives for responsible AI development enough?}

If markets are functioning well and companies and consumers have perfect information about the expected harm of a product, companies should invest the socially optimal amount into product safety \citep{daughety2018market}. In real-world scenarios in which markets may not function perfectly and information asymmetries exist, incentives for companies to invest sufficiently in product safety typically come from three sources: market forces, liability law, and industry or government regulation.\footnote{Other mechanisms include no fault liability systems like mandatory insurance and increasing the information available to consumers \citep{cornell1976safety}.} These three sources of incentives may not provide strong enough incentives for AI companies to engage in responsible AI development, however. We will briefly survey some reasons for this.

\subsubsection{Limited consumer information}

Consumers of AI systems include individuals, private companies, and public institutions. Although different consumers will have access to different levels of information about AI systems, information about the expected harm of AI systems is likely to be quite limited on average. As cutting-edge AI systems become more complex, it will be difficult for consumers not involved in the development of those systems to get accurate information about how safe the systems are. Consumers cannot directly evaluate the safety of aviation software, for example, and will face similar difficulties when it comes to directly evaluating the safety of complex machine learning models. This is compounded by the fact that it is notoriously difficult to explain the decisions made by neural networks \citep{doshi2017towards, olah2018building}. If consumers cannot assess how risky a given AI system is, they cannot adjust their willingness to pay for it accordingly. They are also less able to identify and exert pressure on AI companies that are investing too little in safety \citep{anton2004incentives}.

Consumers could get information about how safe an AI system is by tracking safety failures after its release, but such a `wait and see' strategy could leave both consumers and the public vulnerable to harmful safety failures. This is of particular concern if those safety failures could be irreversible or catastrophic. And the probability of irreversible or catastrophic safety failures is likely to increase as AI systems become more capable and general, since more advanced systems are more likely to be relied upon across a wider range of domains and in domains where failures are more harmful.\footnote{\label{tail}Such safety failures could occur if AI systems have some critical function like controlling national power grids or nuclear weapons systems, or if they can be used to undermine these systems. The more consequential a given technology is, the higher the potential cost of releasing an insufficiently safe version of that technology to both the company and to society. But is worth noting that if the expected cost of catastrophic safety failures is capped by a company's ability to pay then we might expect companies to under-weight these tail-end risks. For a taxonomy of AI risks, see \cite{yampolskiy2015taxonomy}.}

\subsubsection{Limited company and regulator information}

Measuring the safety, security, and social impact of AI systems may turn out to be extremely difficult even for those who understand the technical details of the system. Neural networks are difficult to interpret and as such, failures may be difficult to predict. 
If AI companies are over-confident that their system is not risky, they may under-invest in important risk-reducing measures during development or release a system that causes unintended harm. 

If regulators cannot assess how risky a given AI system is, they may be overly stringent or overly liberal when using regulatory controls \citep{shavell1984liability}. The ability to get accurate information about AI systems therefore seems to be crucial for ex ante safety measures.\footnote{\cite{leike2017ai} introduce simple environments for evaluating the safety of AI agents, and note that future versions of these environments could be used to benchmark the safety performance of AI agents.} 

Our current capacities to identify and measure the expected harms of particular AI systems are extremely limited. We still do not fully understand the decisions made by complex machine learning models  \citep{olah2018building, hohman2018visual} and the high-dimensionality of the inputs to AI systems makes it such that exhaustive enumeration of all possible inputs and outputs is typically infeasible. There may therefore be little consensus about whether a particular system is likely to be unsafe, unsecure, or socially harmful at present. Given this, it is likely that additional capacity will need to be invested by companies or regulators or both in order to decrease these information asymmetries.

\subsubsection{Negative externalities from AI}

Harms caused by AI systems are likely to affect third parties. Biases in algorithmic pre-trial risk assessment are more likely to harm those accused of crimes than those that purchase the tools,\footnote{See \cite{IfAPreTr76:online} on how bias in ML systems could harm those in the California criminal justice system. \cite{pleiss2017fairness} demonstrates the unique difficulties of designing bias-free ML systems.} those that benefit from AI automation may be quite distinct from the people who are displaced by automation,\footnote{\cite{Segal2018} notes that the jobs that have declined as a result of automation so far are intermediate-skill jobs like farming. Although it displaces some workers, automation has positive effects like increased productivity \citep{acemoglu2018artificial}. The overall effect that AI automation will have on the labor force is unclear.} and a major AI disaster---such as an AI system with a faulty reward function\footnote{See \cite{Specific26:online} and \cite{FaultyRe78:online} for examples of faulty reward functions in ML systems.} being integrated into a critical system---could affect a large portion of society that is distinct from the AI company and its consumers. AI also has the potential to be a general purpose technology---a technology that radically affects many sectors of the economy---and if this is the case we should expect its impact to be systemic \citep{ brynjolfsson2018artificial, cockburn2018impact}.

The harms from AI systems may also be difficult to internalize. For example, the social harms that result from an increased use of AI systems---such as reduced trust in online sources---could be complex and diffuse, and it may be difficult to hold any one company strictly liable for them. If the harm is sufficiently large, it may also be too large for a company or insurer to cover all losses (see note \ref{tail}). Finally, AI systems could create negative externalities for future generations that are not in a position to penalize companies or prevent them from occurring \citep{lazear1983intergenerational}. We should expect AI companies to under-invest in measures that could prevent these kinds of negative externalities.\footnote{\label{widespread}Safety failures that affect a large portion of the population will not be treated as externalities by companies if they harm the company or its consumers, though they could still be given insufficient weight.}

\subsubsection{The difficulty of constructing effective AI regulation}

There is currently little in the way of AI-targeted regulation, including government regulation, industry self-regulation, international standards, and clarity on how existing laws will be applied to AI (see note \ref{fn-liability}). Well-designed regulatory mechanisms can incentivize companies to invest appropriate resources in safety, security, and impact evaluation when market failures or coordination failures have weakened the other incentives to do so. Poorly-designed regulation can be harmful rather than helpful, however. Such regulation can discourage innovation \citep{heyes2009environmental} and even increase risks to the public \citep{latin1988good}. 

AI regulation seems particularly tricky to get right, as it would require a detailed understanding of the technology on the part of regulators.\footnote{\cite{hadfield2017rules} and \cite{hadfield2019safeml} outline a regulatory framework for AI that could overcome barriers like information asymmetries and slow response times: key problems for the regulation of new technology.}
The fact that private AI companies can generally relocate easily also means that any attempt to regulate AI nationally could result in international regulatory competition rather than an increase in responsible development.\footnote{\cite{esty2001regulatory} offer an overview of different perspectives on regulatory competition, while \cite{genschel1997regulatory} note that regulatory competition and international co-operation can actually increase levels of regulation. \cite{erdelyi2018regulating} argue that an international AI regulatory agency should be established, but on the grounds that AI has externalities that cross national boundaries.} Regulation that is reactive and slow may also be insufficient to deal with the challenges raised by AI systems. AI systems can operate much faster than humans, which can lead to what \cite{johnson2013abrupt} call `ultrafast extreme events' (UEEs) such as flash crashes caused by algorithmic trading.\footnote{For more on this problem, see \cite{viewpointarticle}. Anticipating, preventing, and responding to catastrophic AI-caused UEEs may present a key challenge in AI safety and policy.}

\subsubsection{The potential for rapid AI development}
 
Some have hypothesized that progress in AI development will be discontinuous (see note \ref{fn-discont}).
On this view, there are some types of AI systems---typically advanced `general' AI systems that are capable of learning effectively across a wide variety of domains---that, if developed, would represent a sudden shift from everything that came before them, and could produce the equivalent of many years of prior progress on some relevant metric.
\footnote{See \cite{Ehrnberg1995} on definitions of technological discontinuities. The AI discontinuity hypothesis should not be confused with the claim that there will be rapid AI development in the future---progress in AI development could be continuous but extremely rapid, e.g. hyperbolic \citep{Hyperbol20:online}---but that there will be a system that represents a sudden leap forward in AI capabilities. It may be possible to achieve a decisive advantage over competitors if progress is rapid but not discontinuous.} If AI progress is discontinuous then developing an AI system that constitutes a sudden leap forward could give a company a large advantage over others, since the next best system would be years behind it in terms of prior progress in the field.\footnote{\cite{bostrom2017superintelligence} claims that such an AI could give a company a `decisive strategic advantage': `a level of technological and other advantages sufficient to enable it to achieve complete world domination' (p.96, \textit{ibid.}). But the concerns we raise here apply even if the advantage is extreme but not decisive in this sense.} Consider the advantage that a company today would gain if they managed to develop something over a decade ahead of current systems used for cyber offense and defense, for example. 

If progress in AI development is discontinuous then market forces and liability law may do little to encourage safe development.\footnote{This may be true even if progress in AI development is continuous but rapid. Even if no single company has a profound advantage over others, mechanisms like regulation and liability could be too slow to catch up with the rate of technological progress. It is worth noting that if AI progress takes this shape then responsible AI development may be more like a one-shot game than an iterated game, which reduces developers' incentives to cooperate on responsible development for reasons that we discuss in the next section.} The value of developing a system that gives a company a huge advantage---that could be used to undermine competition or seize resources, for example---would be largely divorced from the process of getting market feedback. And a company can only be held liable for accidents if these accidents are not catastrophic and the existing legal framework can both keep up with the rapidity of technological progress and enforce judgments against companies. 
Therefore if AI progress is discontinuous, ex ante safety measures like industry self-regulation or international oversight may be more effective than ex post safety measures like market response and liability.
\subsection{Summary}

Incentives to develop safe products generally come from the market, liability laws, and regulation \citep{rubin2011markets}, as well as factors that motivate AI companies beside profits, such as a general desire to avoid doing harm. 
For AI companies, the profit motive to develop AI responsibly is likely to come from the additional revenue generated by AI systems that are more valuable to consumers, the avoidance of reputational harm from safety failures, the avoidance of widespread harms caused by AI systems (see note \ref{widespread}), and the avoidance of tort litigation or regulatory penalties. 

 A key factor that can influence the cost-benefit ratio of responsible AI development that we have not discussed, however, is the competitive environment in which the AI systems in question are being developed. In the next section we will explore the impact that competition between AI companies can have on the incentives that each company has to invest or fail to invest in responsible development.

\section{The need for collective action on responsible AI development}\label{section2}

   We have argued that safer, more secure, and more socially valuable AI systems will tend to have a higher market value, be less likely to cause costly accidents that the company is held liable for, and so on. This means that if a company is guaranteed to be the first to develop a system of this type, we can expect that they will invest resources to ensure that their system is safe, secure, and socially beneficial to the extent that this is incentivized by regulators, liability law, and market forces. This means the more that positive and negative externalities of AI systems have been internalized via these mechanisms, the more that companies can expect to invest in responsible development.\footnote{If the first company could prevent future competitors from entering the market (i.e. the first company could expect to be the only company), it is likely this would reduce but not eliminate market incentives to invest in responsible development \citep{sheshinski1976price}.} 
   
   In this section we will argue that, even with these incentives in place, competitive pressures can cause AI companies to invest less in responsible development than they otherwise would. Responsible AI development can therefore take the form of a collective action problem. We then identify and discuss five key factors that improve the prospects for cooperation between AI companies that could find themselves in a collective action problem over responsible development. 
 
 \subsection{How competitive pressures can lead to collective action problems \label{2.1}}

 To see how the competitive environment could affect investment in responsible development, suppose that several AI companies are working on a similar type of system. If there is a large degree of substitutability between the inputs of different aspects of development, we should not expect AI companies to invest in responsible development beyond the point at which the expected marginal return is lower than the expected marginal return from investing in other areas of development. Suppose each company places less value on coming second than on coming first, less value in coming third than in coming second, and so on. These companies will likely engage in a technological race: a competition to develop a technology in which the largest reward goes to the first company \citep{grossman1985dynamic}.\footnote{As we noted in the previous section, this is a non-trivial assumption that will not hold in all cases.} The resulting dynamics may be similar to those we would expect to see in patent races between firms.\footnote{Patent races have positive effects on innovation, though at the cost of duplicating efforts \citep{judd2012optimal}.} 

There are various strategies companies could use in a ``winner takes more'' race: they could try to develop and maintain a strong technical lead or they could try to to maintain a close position behind the technical leader, for example.\footnote{The best strategy may depend on the competitive environment. \cite{dasgupta1980uncertainty} argue that monopolist companies will attempt to outspend their rivals on R\&D to prevent a duopoly, while \cite{doraszelski2003r} shows that there are conditions in which companies that are behind will invest to catch up.} 
For now, we will assume that the best strategy involves trying to develop and maintain a strong technical lead throughout the race.

Since speed is more valuable when racing against others, we should expect investment into responsible development to be lower when companies are racing against each other.\footnote{How much lower will depend on various features of the race, such as how close it is and the value placed on each position. Note that this argument assumes that investments with even worse expected marginal returns have already been cut. It also assumes that investments in responsible development contribute less to development speed than other available investments: not that they contribute nothing to development speed.} 
  \cite{armstrong2016racing} point out that in an AI development race, responsible development could be prey to a ``race to the bottom'' dynamic.
 Consider what happens if one company decides to increase their development speed by decreasing their investment in safety, security, and impact evaluation. This increases their expected ranking in the race and decreases the expected ranking of others in the race. A decrease in expected ranking gives competing AI companies an incentive to decrease their own investment in these areas in order to maintain or increase their expected ranking in the race.\footnote{In this scenario, companies have full information about the investments made by other companies are their likelihood of winning. But this assumption is not necessary, since companies can invest in accordance with their expectation about the investments and win probabilities of other companies. \cite{armstrong2016racing} explore scenarios in which AI companies have different levels of information about their own and others' capabilities.} 
 
 We might ask why racing to the bottom on product safety is not ubiquitous in other industries in which decreasing time-to-market is valuable, such as in the pharmaceutical industry.\footnote{It is worth noting that similar concerns about the desire to develop quickly conflicting with risk management have been expressed in other industries that involve novel technology, such as the use of nanoparticles and nanotechnology in the food industry \citep{morgan2005development, cushen2012nanotechnologies}.} 
 The most plausible explanation of this difference is that the cost of safety failures has been internalized to a greater extent in more established industries via external regulation, self-regulation, liability, and market forces.
 These mechanisms can jointly raise the ``bottom'' on product safety to a level that is generally considered acceptable by regulators and consumers.\footnote{ It could also be the that the best strategies in technological races do not involve trying to develop a strong technological lead, or that there are unidentified factors that make racing to the bottom on product safety undesirable: factors that may apply equally to the development of AI systems.}

In a race to the bottom on safety, competing AI companies could reduce their investment in responsible development to the point that winning the technology race---successfully developing the system they are racing to develop before others---is barely of net positive value for the winner even after all the first-mover advantages, including positive reputational effects, the ability to capture resources like data, hardware and talent, and creating switching costs for consumers, have been taken into account.\footnote{The company with the winning system could even consider their own system to be worse than developing nothing at all absent competition, though this would only happen if they considered the release of the alternative winning system to be even worse for them than the release of their own worse-than-nothing system.}

\subsection{When competition has negative rather than positive effects \label{2.2}}

The race to the bottom on safety described above is a collective action problem: a situation in which all agents would be better off if they could all cooperate with one another, but each agent believes it is in their interest to defect rather than cooperate.\footnote{This is a weakening of the definition that Jon Elster derives from \cite{schelling2006micromotives}, which states `First, each individual derives greater benefits under conditions of universal cooperation than he does under conditions of universal noncooperation. Second, each derives more benefits if he abstains from cooperation, regardless of what others do.' \cite[p.139]{elster1985rationality}. We simply replace `regardless of what others do' with `given what we expect others will do.' See \cite{holzinger2003problems} for a broader definition and taxonomy.} As  \citet[p. 78]{doi:10.1177/1043463189001001006} states, ``the inclinations of individuals (that is, each actor's preferences regarding his or her own behavior) are in conflict with regulatory interests (that is, each actor's preferences regarding the behavior of others). The collective action problem arises when a group possesses a common interest, or faces a common fate.''

In a race to the bottom on safety, it is in each company's interest to reduce their investment in responsible development in order to increase development speed. If all companies do this, however, there is a single equilibrium: one in which much or all of the value that could have been be gained with coordination is destroyed. If each company defects, they will have a similar position in the race to the one that they would have had if they had all successfully coordinated, but they will be developing systems that are more risky than the ones they would have developed if they had all managed to successfully coordinate. In other words, the situation in which they find themselves is strictly worse than the situation in which coordination was successful. 

Collective action problems between companies can have positive effects on consumers and the public. A price war is a collective action problem between companies with mostly positive effect on consumers, for example, as it results in lower prices.\footnote{When companies engage in price wars, prices often end up close to their marginal cost of production \citep{bresnahan1987competition}. Prices can even be temporarily set below the marginal cost of production in order to push competitors out of the market \citep{guiltinan1996aggressive}, sometimes in violation of antitrust.} Antitrust law exists to maintain competition between companies that has a positive effect on consumers and to prevent collusion between companies that has a negative effect on consumers (e.g. price fixing).

When there are negative effects from production that are not captured by the incentives facing producers (i.e. negative externalities), however, competition does not lead to the socially optimal outcome. If this outcome is also bad for the producers, it is a collective action problem for producers.

A race to the bottom on safety falls into this category if it results in AI systems with safety levels below what is socially optimal and below what AI companies would prefer. Pollution by companies is another example of a collective action problem between companies that has a negative effect on the public \citep{leveque1999externalities}.

Before discussing strategies for cooperation such as self-regulation in more depth, however, it will be useful to understand the incentives that AI companies have to abide by norms that involve mutual investment in responsible AI development. This will be the focus of the remainder of this section.
\subsection{Incentives to cooperate in collective action problems \label{2.3}}

In an AI development race, companies ``cooperate'' if they maintain some acceptable level of investment in responsible development and they ``defect'' if they fail to maintain this level of investment, thereby acting in their own interest (hypothetically) and against the collective interest. Encouraging companies to cooperate should therefore not be confused with encouraging them to stop competing. Companies agreeing not to compete across the investment in safety dimension does not imply that they will cease to compete across the R\&D dimension. Competitive dynamics that contain cooperative elements are sometimes referred to as a ``coopetition''.\footnote{See \cite{bengtsson2000coopetition} and \cite{tsai2002social}.}

If companies have incentives to prevent or mitigate collective action problems that have negative effects on consumers or the public then we should expect the companies themselves (and not just third parties like government regulators) to take steps to solve them. And companies often do attempt to cooperate to prevent or solve collective action problems of this sort. One example of a mechanism used to this end is industry self-regulation. \citep{gunningham1997industry}.\footnote{Industry self-regulation can also be incentivized by government regulators via meta-regulation \citep{parker2007meta, coglianese2010meta}.} Examples of self-regulation include Responsible Care: a self-regulation program in the US chemicals industry \citep{gamper2013does}, and the Institute of Nuclear Power Operations (INPO): an industry organization that conducts inspections and facilitates the sharing of best practices in the nuclear power industry \citep{davis2012deregulation, hausman2014corporate}.\footnote{Other examples of self-regulation can be found in a variety of industries, as self-regulation is sometimes used to preempt government regulation \citep{lenox2007prospects}. How successful such self-regulation is at reducing the negative effects of collective action problems varies a great deal by industry. The INPO is generally considered to be a more successful self-regulatory scheme than Responsible Care, for example \cite[pp. 126-7]{cohen2015self}. This may be because the INPO, unlike Responsible Care, has an agreement with a government regulator, the Nuclear Regulatory Commission, which can monitor the program and provide meaningful sanctions, which may be required for successful self-regulation \citep{king2000industry}. \cite{cullen1} explores one possible form of antitrust-compliant self-regulation in the AI industry.} 

In order to identify features that affect the degree to which it is in a company's interest to cooperate on responsible development, it will be helpful to highlight features that increase incentives to cooperate in collective action problems generally. To do this, consider the payoff matrix of a cooperate-defect game in which two agents (AI companies) can cooperate (develop responsibly) or defect (fail to develop responsibly). Here the first letter in each pair represent the expected payoff for Agent 1, and the second letter in each pair represents the payoff for Agent 2.\footnote{We assume that these expected utilities have already factored in agents' attitudes towards risk and discuss some of the simplifications of this framework below.}

\FloatBarrier
  \begin{table}[h]
  \centering
    \setlength{\extrarowheight}{2pt}
    \begin{tabular}{cc|c|c|}
      & \multicolumn{1}{c}{} & \multicolumn{2}{c}{Agent 2}\\
      & \multicolumn{1}{c}{} & \multicolumn{1}{c}{Cooperate}  & \multicolumn{1}{c}{Defect} \\\cline{3-4}
      \multirow{2}*{Agent 1}  & Cooperate & $a_1, a_2$ & $b_1, b_2$ \\\cline{3-4}
      & Defect & $c_1, c_2$ & $d_1, d_2$ \\\cline{3-4}
    \end{tabular}
    \vspace{0.4cm}
    \caption{A Normal Form Cooperate-Defect Game}
  \end{table}
  \FloatBarrier
  
Let $p$ be the probability that Agent 1 assigns to  Agent 2 cooperating and let $q$ be the probability that Agent 2 assigns to  Agent 1 cooperating. 
We assume it is rational for Agent 1 to cooperate if the expected value of cooperation (the likelihood Agent 2 will cooperate times $a_1$ plus the likelihood Agent 2 will defect times $b_1$) is greater than the expected value of defection (the likelihood Agent 2 will cooperate times $c_1$ plus the likelihood Agent 2 will defect times $d_1$). We assume the same is true of Agent 2.\footnote{In other words, it is rational for Agent 1 to cooperate if $p\times a_1 + (1-p)\times b_1 > p\times c_1 + (1-p)\times d_1$ and it is rational for Agent 2 to cooperate if  $q \times  a_2 + (1-q)\times c_2 > q \times b_2 + (1-q) \times d_2$. These two agents are in a collective action problem if it is irrational for both agents to cooperate, 
but $a_1>d_1$ and $a_2>d_2$. If both sides of these equations are equal then defecting and cooperating are both rationally permissible for the agent. Note that if $a_1 + a_2 > d_1 + d_2$ but $a_1 \ngtr d_1$ or $a_2 \ngtr d_2$ (i.e. defecting is rational for at least one agent but mutual cooperation creates more total value for both agents than mutual defection does) then the likelihood of cooperation increases if redistribution is possible, i.e. if the agents can bargain towards a solution.} 
This lets us identify five highly interrelated factors that increase an agent's incentive to cooperate.
These factors are as follows, where expected values are relative to the agent's beliefs:\footnote{\label{five}If Agent 1 and Agent 2 are not in an anti-coordination game 
then Agent 1's incentives to cooperate increase as (1) $p$ increases, (2) the expected value of $a_1$ increases, (3) the expected value of $c_1$ increases, (4) the expected value of $b_1$ decreases, and (5) the expected value of $d_1$ decreases. 
Naturally, the inverse of each of these factors will decrease the agent's incentive to cooperate.} 

\begin{itemize}
   \item[] \hspace{-0.6cm} \textit{(1) High Trust:} being more confident that others will cooperate ($p, q$)\footnote{This is not the only definition of `trust', but it is the one that is most relevant to the current analysis.} 
    \item[] \hspace{-0.6cm} \textit{(2) Shared Upside:} assigning a higher expected value to mutual cooperation ($a_1$, $a_2$)
    \item[] \hspace{-0.6cm} \textit{(3) Low Exposure:} assigning a lower expected cost to unreciprocated cooperation ($b_1$, $c_2$)
    \item[] \hspace{-0.6cm} \textit{(4) Low Advantage:} assigning a lower expected value to not reciprocating cooperation ($c_1$, $b_2$)
    \item[] \hspace{-0.6cm} \textit{(5) Shared Downside} assigning a lower expected value to mutual defection ($d_1$, $d_2$)
\end{itemize}

The last four factors each refer to the expected value of an action conditional on the behavior of the other agent, such as cooperating and having your cooperation reciprocated. 
Note that the expected value of an action depends on how good the agent perceives the outcome to be and how likely the agent perceives it to be. This means that an agent could be in a `low exposure' scenario if she considers unreciprocated cooperation to be not very valuable or not very likely or both.
We can provide agents with evidence about the likelihood and value of each outcome by changing the world in some perceptible way, e.g. by offering a reward for responsible development, or by giving them evidence about the way the world already is, e.g. by correcting false beliefs.

It is useful to separate the degree of trust (factor 1) from incentives (factors 2-5) in order to discuss its role in cooperation, but trust is not independent of incentives or vice versa.
If one agent comes to trust an agent more, this increases the expected value of the outcomes that involve cooperation.\footnote{We say `in the situations we consider here' because if  the agents are in an anti-coordination game then increasing Agent 1's trust in Agent 2 will decrease Agent 1's incentives to cooperate.} The same is true in reverse: if the expected value of the outcomes that involve cooperation increase, it is more likely that the other agent will cooperate. In other words, increasing trust can increase incentives to cooperate, and increasing incentives to cooperate can increase trust between agents.\footnote{Again, this will not be true in certain anti-coordination games.}

This means that if a company can provide information about itself that increases the probability the other assigns to it cooperating, this will increase the degrees of trust between the companies and make it more likely each company's trust threshold will be met. Two important facts follow from this. First, information that companies provide about their intentions and actions---how transparent they are---can play an important role in whether other companies will cooperate with them. Second, trust is prey to virtuous and vicious cycles. If one company demonstrably increases its trust in another, the other company should increase its trust in return. But if one company demonstrably decreases its trust in another, the other company should decrease its trust in return.\footnote{This is one reason why a degree of `forgiveness' can be strategically valuable: it can prevent errors, misinterpretations, or aberrant behavior from plunging both players into a vicious cycle of distrust prematurely, and can pull players out of such a cycle \citep{axelrod1980more}}.

A real world race to the bottom on safety would unfold over many interactions. 
The factors identified here also increase the prospect of cooperation in sequential games, however.\footnote{The main adjustment we need to make to the factors above in extensive form games will be to the first factor: high trust. If we let $C_i$ mean that agent $i$ cooperates and assume that agents can only either cooperate or defect, in extensive form games our `high trust' factor would say that the incentives for Agent 1 to cooperate with Agent 2 increase as $p(C_2|C_1)$ and $p(\neg C_2|\neg C_1)$ increase. The other four factors can remain largely unchanged.} 
And iterated collective action problems are generally easier to solve than one-shot collective action problems because, in iterated collective action problems, players have an incentive (and opportunity) to cooperate early in the game in order to establish trust and avoid retaliation.\footnote{In an iterated Prisoner's Dilemma, for example, cooperation can be incentivized by things like the threat of retaliation and the promise of reciprocity \citep{axelrod1984evolution, nowak2006five}. The promise of reciprocity increases the expected value of mutual cooperation today (shared upside) and the threat of retaliation decreases the expected value of betraying the cooperation of others today (low advantage). The payoff structure of the iterated Prisoner's Dilemma may therefore be more like that of a Stag Hunt \cite[p.123]{seabright1993managing}. See \cite{mailath1991extensive} on the extent to which important features of extensive form games can be preserved in normal form.} Using one-shot games to illustrate our points is therefore more likely to skew us towards undue pessimism about our ability to solve races to the bottom rather than undue optimism.

One shortcoming of our analysis, however, is that it appeals to an overly simplified conception of cooperation and defection. For example, we assume that the options available to agents can be divided into `cooperation' and `defection'. In reality, cooperation will come in varying degrees---companies can invest different amounts in responsible development, for example---and it would be better to talk about the degree of cooperation that we can expect between agents.\footnote{We are attempting to illustrate the general structure of reasons to cooperate in AI development rather than analyzing a particular case in detail.} We also assume that companies will make an intentional decision to coooperate or defect over time. In reality, companies could fail to foresee the consequences of investing very little into areas like safety, and may therefore defect without intending to. Third, we assume that both companies perfectly understand the actions and assertions of the other. In reality, it may not be clear whether a company is living up to an agreement to develop AI responsibly. If agreements are not clear then there may not be a bright line between defection and non-defection that companies can respond to \citep{chassang2010building, gibbons2012managers}. A more complete analysis of collective action problems in AI development should build a more realistic model of what cooperating and defecting during AI development would look like.

We have argued that in order to ``solve'' a collective action problem, we can try to transform it into a situation in which mutual cooperation is rational. If we can transform it into a situation in which agents have lower minimum trust thresholds (generally determined by the payoff matrix)
 and greater trust of each other---greater confidence that if they cooperate, others will reciprocate \cite[p. 9]{kydd2007trust}---then we should expect a higher degree of mutual cooperation.
 \footnote{
 Sometimes collective action problems are the result of one or more agents having mistaken beliefs about the expected value of cooperating and defecting. When this is the source of the problem, it can be `solved' by correcting these misconceptions.}
 Given this, we should expect ``lower conflict'' collective action problems---problems in which agents have stronger incentives to cooperate---to be easier to solve than ``higher conflict'' collective action problems---problems in which agents have weaker incentives to cooperate.\footnote{
 Scenarios in which agents have stronger incentives to cooperate with one another involve less `conflict' \citep{robinson2005conflict, schelling1980strategy}. How easy it is to solve collective action problems depends both on the degree of conflict involved and the nature and magnitude of the resources we have at our disposal. 
 For example, the Stag Hunt is easier to solve than the Prisoner's Dilemma. All possible adjustments to the Prisoner's Dilemma that result in a solution will, if applied to the Stag Hunt, result in a solution to this problem as well. But only some adjustments to payoffs and probabilities that solve the former would also solve the latter.}

\subsection{The cooperative factors in AI development \label{2.4}}

Whether an AI development race will result in a collective action problem and, if so, how bad it will be are both open questions.\footnote{It is also worth bearing in mind that scenarios can be superficially similar to collective action problems even though it is in everyone's interest to cooperate.}
But there are many features of an AI development race that affect both the likelihood and severity of collective action problems. For example, having close frontrunners would likely worsen a collective action problem---would reduce the tractability of resolving it---because this increases the expected value frontrunners will assign to not reciprocating the cooperation of others (low advantage) and therefore increases the probability they assign to not having their own cooperation reciprocated (high trust).\footnote{This is consistent with the conclusion of \cite{armstrong2016racing} that frontrunners will take more risks if they have a close competitor. The claim that competition is more intense among close competitors has also been made in the literature on R\&D races also \citep{grossman1985dynamic, Harris1987}.} Similarly, a misaligned perception of the risks associated with different AI systems could worsen a collective action problem if it causes less cautious companies to assign a lower cost to not reciprocating cooperation (low advantage), which could increase the probability that cautious companies assign to having their cooperation unreciprocated by less cautious companies (high trust) and increase the expected harm that cautious companies expect to arise from incautious companies getting ahead this way (low exposure). 

Features that affect the likelihood and severity of a collective action problem for responsible development can be used to decrease its likelihood and severity if they are are features that we can control. For example, fundamental distrust between companies is likely to worsen a collective action problem because companies are less likely to expect that their cooperation will be reciprocated (high trust). Building trust between AI companies can therefore decrease the severity of collective action problems. An AI race development in which the expected value of winning is much greater than the expected value of losing is also likely to have a worse collective action problem (low exposure and low advantage).\footnote{We have focused on cases that involve cooperation, but we can use the more cooperation-neutral factors in note \ref{five} to look at the expected cost to one company if another company wins regardless of the degree of cooperation between the two companies. To give another example of this, \cite{armstrong2016racing} discuss the level of enmity between companies. Higher enmity would then be expected to worsen a collective action problem by increasing the cost of losing the race to the other company (low exposure).} If close frontrunners worsen collective action problems, AI companies may agree to take steps to avoid engaging in a harmful race to the bottom on safety. For example, citing concerns about race dynamics, \cite{OpenAICh43:online} have stated that ``if a value-aligned, safety-conscious project comes close to building AGI before we do, we commit to stop competing with and start assisting this project.'' 

The mechanisms to incentivize investment in product safety outlined in the previous section---market forces, regulation, and liability---all operate to prevent collective action problems for product safety. Consumers often pay less for products that are unsafe (low advantage and shared downside) and more for safe products (shared upside and low exposure).  Government regulation either removes the option to underinvest in safety or increases the cost of underinvesting in safety via sanctions and fines (low advantage and shared downside). And the possibility of being held liable for harms caused by unsafe products decreases the expected value of underinvesting in safety to get ahead (low advantage and shared downside).

Market forces, regulation, and liability are all mechanisms operating outside of the AI industry that affect the incentives that AI companies have to develop responsibly. But if responsible AI development is a collective action problem then each AI company expects to benefit from being in a better equilibrium and therefore has an incentive to ensure that the AI industry itself collectively coordinates to maintain some acceptable level of responsible development.  Companies should be willing to invest in cooperative mechanisms to the degree that these mechanisms increase the likelihood that they will be able to capture the cooperation surplus: the additional expected value that cooperation would generate for them.\footnote{This concept is similar to the Harsanyi dividend, which quantifies the value created by a coalition \citep{harsanyi1963simplified}. The value of trust as a commodity is explored by \cite{dasgupta2000trust}. Companies should be willing to pay more for credible demonstrations of their intention to cooperate.}

This means that industry-led mechanisms like greater self-regulation could also be developed to incentivize responsible AI development.\footnote{\cite{king2000industry} highlights the difficulties of self-regulation by looking at the chemical industry's Responsible Care program. A self-regulatory program that is considered more successful, however, is the Institute of Nuclear Power Operations (INPO). See \cite{coglianese2010meta} for an analysis of both.}

\subsection{Summary}

In this section we argued that responsible AI development may take the form of a collective action problem. We also identified five factors that generally increase the likelihood of mutual cooperation and can help solve such collective action problems. In the next section we will translate this into more concrete suggestions for increasing cooperation on safety between AI companies.

\section{Strategies to improve AI industry cooperation on safety}
\label{sec:factors}

In the previous section, we argued five factors make it more likely that AI companies will cooperate if they are faced with a collective action problem: (1) being more confident that others will cooperate, (2) assigning a higher expected value to mutual cooperation, (3) assigning a lower expected cost to unreciprocated cooperation, (4) assigning a lower expected value to not reciprocating cooperation, (5) assigning a lower expected value to mutual defection.

These five factors give high-level direction regarding how to ensure that the fruits of cooperation in AI are realized. However, it is not always obvious what these five factors mean in the real world, so there is a need for translating these factors into tangible policy strategies that various actors can implement in order to improve cooperation prospects. 

It is impossible to prescribe such strategies fully in advance, because we lack information about the future which would be needed in order to make informed future decisions, and because a particular policy proposal could be effective if well-implemented but counterproductive if poorly executed. However, while detailed, long-term policy prescriptions would be premature today, there are several coarse-grained strategies that seem robustly desirable even if some of the low-level details require research, dialogue, and passage of time before they can be clarified. 

We believe that the four strategies we identify in this section are robustly desirable in the sense that they all have substantial benefits with respect to at least one of the factors above, and are unlikely to be very harmful with respect to the others. 
\subsection{Promote accurate beliefs about the opportunities for cooperation}

As noted in prior sections, there are multiple competing conceptions of AI development. In cases where people are demonstrably uninformed about key aspects of AI development, it is likely beneficial to correct them, and more generally for stakeholders to make nuanced public statements consistent with the spirit of AI development that involves cooperation on norms of responsible development. 

Some misconceptions that should be corrected in order to improve prospects for such cooperation include incorrect beliefs that safety and security risks can be safely ignored \citep*{2018arXiv180207228B, amodei2016concrete, Building2:online}, an unwarranted focus on relative gains and losses instead of absolute gains and losses (shared upside, low exposure, low advantage, shared downside), and mistaken belief in interests being more misaligned than they are (low exposure and low advantage), In addition to correcting specific misconceptions, there is also likely value in proactively informing people about the case for cooperating on responsible development generally. 

For example, recent years have seen substantial effort by researchers and activists to highlight the biases being learned by deployed AI systems in critical societal domains such as criminal justice and in widely used technological platforms such as recommender systems. This work has highlighted the risks of incautious development to a large and growing swathe of the AI community. Similarly, concerns have been raised about both the bias, efficacy, and other properties of medical AI systems, as well as self-driving vehicles and other emerging technologies. Analyzing and communicating these sorts of risks is critical for generating interest in cooperation among a sufficiently wide range of actors, as well as in identifying appropriate norms around research, publication, and deployment given the safety risks and the ways of mitigating them that have been identified.  

In many cases, common knowledge that multiple parties share a concern or interest can be critical for the initiation of cooperation, and a misconception that parties lack such a shared concern or interest could be damaging to cooperation on issues like safety. Avoiding such misunderstanding may be particularly important in the case of international cooperation on responsible AI development across distinct countries with different languages and cultural frames of reference.

Propagating accurate information about existing beliefs can also be valuable, as it allows multiple parties to stabilize their expectations. For example, the Asilomar AI Principles \citep{AIPrinci58} commit the many signatories to arms race avoidance, and various statements of principles before and after this have similarly committed many actors to various (admittedly still abstract) cooperative statements and actions. Expanding the breadth and depth of such dialogue, especially across cultural and language boundaries, will be critical in fostering understanding of the large gains from mutual responsible development (shared upside) and the large losses from mutual irresponsible development (shared downside), and in establishing common knowledge that such understanding exists (high trust).

It is possible to create positive spirals of trust, in which an increase in one party's trust causes the trusted party to increase their trust in turn. We can also stumble into negative trust spirals, however, in which a loss of trust leads to further distrust between parties. It is therefore also important to avoid feeding into unnecessarily adversarial rhetoric about AI development, lest it become self-fulfilling \citep{kreps}.

\subsection{Collaborate on shared research and engineering challenges}

On a range of possible research challenges---from basic AI research to applied AI projects to AI safety and security research---it can be beneficial for multiple parties to actively pool resources and ideas, provided this can be done in a way that is procompetive and compliant with antitrust laws \citetalias{ftc-collab}, 
does not raise security concerns for the companies participating, and so on.

Joint research can provide value for cooperation via useful technical insights (such as solutions to safety problems; low exposure and low advantage), stabilizing expectations regarding who is working on what via public information about joint investments as well as interpersonal dialogue (versus work being more shrouded in secrecy; high trust and shared downside), concretizing the joint upsides of AI (e.g. AI for good collaborations; shared upside), and facilitating more societally beneficial publication and deployment decisions by various actors (e.g. via collaborative analysis of the risks of specific systems; shared upside).\footnote{For example, OpenAI’s approach to the release of the GPT-2 language model family \citep*{radford2019language} involves staged release, in which a model is released incrementally due to safety and security concerns, and partnership-based sharing, in which a model is shared with a small number of research partners to enable research on that system without necessarily requiring broad-based access. This experiment in responsible publication, and others like it such as the Allen Institute for Artificial Intelligence and the University of Washington’s approach on their Grover family of language models, may help to ``derisk'' this particular form of research-level collaboration discussed in the next subsection.} Note that we refer specifically here to active and explicit research collaboration, of which some already occurs, alongside a much greater amount of implicit collaboration on AI research that already exists due to the high degree of openness in the AI research community. 

Active and explicit research collaboration in AI, especially across institutional and national borders, is currently fairly limited in quantity, scale, and scope. This is for a range of reasons. In order to maintain legitimate academic and industrial competition, researchers or their managers may be averse to publishing certain research outputs early or at all. And research ideas, results, datasets, and code can be hard to disentangle from proprietary product plans and technical infrastructure. Furthermore, safety or security considerations can in some cases make the joint analysis of a particular system more challenging than it would otherwise be \citep*{radford2019language}. There are also linguistic and logistical barriers to collaborating across long distances and across different cultures and languages. 

While we acknowledge that such challenges exist, we advocate a more thorough mapping of possible collaborations across organizational and national borders, with particular attention to research and engineering challenges whose solutions might be of wide utility. Areas to consider might include joint research into the formal verification of AI systems' capabilities and other aspects of AI safety and security with wide application; various applied ``AI for good'' projects whose results might have wide-ranging and largely positive applications (e.g. in domains like sustainability and health); coordinating on the use of particular benchmarks; joint creation and sharing of datasets that aid in safety research; and joint development of countermeasures against global AI-related threats such as the misuse of synthetic media generation online. 

\subsection{Open up more aspects of AI development to appropriate oversight and feedback}

Openness about one's beliefs, actions, and plans is critical to establishing trust generally. In the case of AI development, those building and deploying AI systems need to provide information about their development process so that users can make informed decisions. Likewise, governments need to be able to appropriately oversee safety-critical AI systems, and (in the absence of relevant regulation) companies need to be able to provide information to one another that shows they are following appropriate norms. 

The general appeal of openness for cooperation-related reasons does not imply that all aspects of AI development should always be open, and as AI systems become more capable, it will be increasingly important to decide responsibly what should and shouldn't be made open  \citep*{2018arXiv180207228B, bostrom2017strategic, ClopenAI24:online}. Full transparency is problematic as an ideal to strive for, in that it is neither necessary nor sufficient for achieving accountability in all cases \citep*{desai2017trust, ananny2018seeing}. Further, some information about AI development cannot or should not be shared for reasons of safety, security, ethics, or law. For example, AI developers might legitimately be wary of releasing code that is intimately tied to proprietary infrastructure, and should certainly be wary of releasing private data as well as AI systems that are easily amenable to misuse. 

Given that full openness is rarely called for, but that some openness is required for building trust, there is a need for continuing effort to implement existing modes of trust-building in AI, as well as to discover new ones. Different mechanisms for achieving openness regarding how AI systems are developed and operated include, e.g., publicizing decision-making principles and processes, explaining publication/release decisions, sharing accessible information about how particular AI systems and broad classes of AI systems work, allowing external visitors to the lab, and opening up individual AI systems to detailed scrutiny (e.g. via bug bounties or open sourcing). 

Such openness is critical in allowing reputation to play its stabilizing role in cooperation. Indeed, some actors have explicitly pointed to the challenges of monitoring the development and use of lethal autonomous weapons as as a reason not to agree to strict rules, suggesting that the inability to track others' behavior reliably could be a bottleneck on some forms of mutually beneficial cooperation (e.g. joint restraints on weapons development). In cases such as this, a richer set of tools for opening up actors to critical scrutiny and feedback (while managing the associated risks) would be useful, and we encourage continued exploration of approaches such as those mentioned above as well as others in order to widen the range of cooperative actions available to AI developers. 

In combination, the appropriate application of transparency mechanisms such as these should reduce the severity of concerns about others behaving irresponsibly (low exposure), reduce the temptation to defect in partially competitive situations (low advantage), and increase confidence that others’ statements about their behavior are accurate (high trust). Openness is a particularly powerful strategy, and applicable to a wider range of cooperation problems, if it can be gradually ratcheted up in an iterative fashion between parties, as opposed to happening all at once. This gradual approach can reduce the temptation to defect at any particular stage (low advantage) and increase confidence in others cooperating (shared downside).

\subsection{Incentivize adherence to high standards of safety}

Cooperative actors might want to introduce additional incentives (reward and/or punishment) related to responsible AI development beyond those that exist today, or would exist by default in the future. E.g. such actors might strongly value compliance with certain norms intrinsically, and prefer that those who comply with appropriate norms be rewarded; or one might want to deliberately bring about an incentive for oneself to act in a certain way, as a commitment mechanism; one might also want to use incentives as a complement to other governance tools such as monitoring of behavior and direct regulation; and one might want to generally influence the incentives of many actors in a particular direction, and support policies that bring this about.

There are several categories of incentives that one might want to consider in this context. Creating incentives for key actors to act cooperatively, if done effectively, would help with all five factors simultaneously. Potential incentives include:

\begin{itemize}
  \item Social incentives (e.g. valorizing or criticizing certain behaviors related to AI development) can influence different companies' perceptions of risks and opportunities
  \item Economic incentives (induced by governments, philanthropists, industry, or consumer behavior) can increase the share of high-value AI systems in particular markets or more generally, and increase attention to particular norms\footnote{
Mutual agreements to distribute the economic gains from winning the AI development \citep{cullen2} could also decrease the severity of collective action problems.}
  \item Legal incentives (i.e. proscribing certain forms of AI development with financial or greater penalties) could sharply reduce temptation by some actors to defect in certain ways. 
  \item Domain-specific incentives of particular relevance to AI (e.g. early access to the latest generation of computing power) could be used to encourage certain forms of behavior.
\end{itemize}

As argued in each case above, these strategies are robustly desirable from the perspective of enabling cooperation, but our articulation of them leaves many questions unanswered. In particular, sharpening these recommendations and adapting them over time will require technical and social scientific research, creative institutional design, and bold policy experimentation, e.g. via regulatory markets as discussed in \cite{hadfield2019safeml}.

\section{Conclusion and Future Directions}\label{sec:conc}

In this paper we have argued that competition between AI companies could create a collective action problem for responsible AI development. We have identified five key factors that make it more likely that companies will cooperate on responsible development: high trust, shared upside, low exposure, low advantage, and shared downside. We have shown that these five factors can help us to identify strategies to help AI companies develop responsibly and thereby realize the gains from cooperation. This also has important positive externalities for consumers and the general public.

If our analysis is on the right track then it is best thought of as the beginning of a program of research, rather than the last word on the subject. Much work needs to be done to identify whether collective action problems for responsible AI development will occur if we vary who is developing AI, how many entities are developing AI, what systems they are developing, and so on. More work must also be done to identify and evaluate strategies that can prevent or mitigate these kinds of collective action problems across a wide range of possible scenarios. 

The possible future research directions on this issue are broad and we do not aim to provide a comprehensive list of them here, but examples of potentially fruitful research questions include:

\begin{enumerate}
\item How might the competitive dynamics of industry development of AI differ from government-led or government-supported AI development? 
\item What is the proper role of legal institutions, governments, and standardization bodies in resolving collective action problems between companies, particularly if those collective action problems can arise between companies internationally?
\item What further strategies can be discovered or constructed to help prevent collective action problems for responsible AI development from forming, and to help solve such problems if they do arise? What lessons can we draw from history or from contemporary industries?
\item How might competitive dynamics be affected by particular technical developments, or expectations of such developments?
\end{enumerate}

As we noted at the outset, there is substantial uncertainty about the nature and pace of developments in AI. If the impact of AI systems on society is likely to increase, however, then greater attention must be paid to ensuring that the systems being developed and released are safe, secure, and socially beneficial. 
In this paper we argued that existing incentives to develop AI responsibly may be weaker than is ideal, and that this may be compounded by competitive pressure between companies, leading to a collective action problem on the responsible development of AI.

That such collective action problems will arise or that they will be maintained if they do arise is far from a foregone conclusion, however. Finding ways of preventing and solving these problems may require new ways of building trust in novel technological contexts, and in some cases to assume some risk in the expectation that others will reciprocate in turn. While intellectually and politically challenging, we think such efforts are integral to realizing the positive-sum potential of AI.

\section*{Acknowledgments}

We are grateful to Michael Page, Jack Clark, Larissa Schiavo, Carl Shulman, Luke Muehlhauser, Geoffrey Irving, Sarah Kreps, Paul Scharre, Michael Horowitz, Robert Trager, Tamay Besiroglu, Helen Toner, Cullen O’Keefe, Rebecca Crootof, Ben Garfinkel, Adam Gleave, Jasmine Wang, and Toby Shevlane for valuable feedback on earlier versions of this paper.

\newpage

\raggedright
\bibliography{cooperation-in-ai}

\end{document}